# Effect of Composition on Microstructural Evolution during Homogenization of 7XXX Alloys

Pikee Priya[a,#], Yiwei Sun[a], D. R. Johnson[a], K. P. Trumble[a], M. J. M. Krane[a]

[a] Purdue Center of Metal Casting Research, School of Materials Engineering, Purdue University, West Lafayette, IN, USA.

[#]ppriya@iisc.ac.in

## Abstract

The effect of composition on microstructure both at the length scale of the secondary dendrite arm spacing and nano-sized dispersoids during homogenization of Al-Zn-Cu-Mg-Zr alloys has been studied. A comprehensive model that can predict the microstructure at both the length scales has been used for the study. The microstructure predicted has been compared to that for two homogenized samples from a directionally solidified AA7050 sample and a reasonable match has been found. The initial as-cast microstructure for different compositions is calculated using Scheil type solidification from Thermo-Calc$^{TM}$. The initial microstructure has a considerable influence on microstructural evolution during homogenization. To take advantage of the decreased solid solubility of Zr in α-fcc, cooling rates during solidification must be high enough to prevent precipitation of primary $Al_3Zr$. Under solidification with industrial cooling conditions solute rich alloys leads to fewer dispersoids. Based on the study, an improved composition range of 6-8%Zn, 1-2%Cu, 1-2%Mg and 0.1-0.15%Zr for 7XXX alloys has been proposed.

## 1. Introduction

7XXX series alloys which are Al-Zn-Cu-Mg alloys with minor additions of Zr or Sc are used in aerospace applications due to their high strength, fracture toughness and corrosion resistance. Compositional variations have significant effects on the microstructure after casting and homogenization which influences the workability and mechanical properties of these alloys[1]–[5]. The varying amounts of alloying elements lead to stabilization of coarse interdendritic particles[6] as well as affect the distribution of fine dispersoids across the grains[7] during homogenization of the as-cast alloys. Thus, the alloying elements affect the microstructure both at the secondary dendrite arm spacing (SDAS) and the nano-sized dispersoid length scales.

The composition determines the stable or metastable phases which remain after processing. The η, T or S phases that remain affects the mechanical properties: strength, ductility and fracture toughness of the component[3], [5]. While the finely distributed η and η' phases are desirable, T and S are not. η and η' formed during age-hardening of these alloys, increase the strength and fracture toughness of the alloy[3], [5]. T and S are low melting phases [8]which might melt during thermo-mechanical processing. Also, the S phase is brittle and affects the fracture toughness of the material [9], [10].

Increase in the Zn:Mg ratio decreases the amount of T and S phases in the alloy [6]. Increasing the Zn also improves the strength of the component [5]. However, very high amounts of (Zn+Cu+Mg) make the component quench sensitive [6], requiring very high cooling rates to subdue precipitation which may lead to residual stresses[11][12] and deformation which is not uniform. Decreasing the Mg and Cu content decreases the amount of T phase and S phase both of which are undesirable. However, they are needed for age-hardening of these alloys [13]. Solute content in these multi-component Al alloys also affect the solubility of Zr in Al affecting their microsegregation during casting and nucleation/growth rates during homogenization. Because there is a variation of Zr composition across the SDAS which is not eliminated after homogenization due to its low diffusivity, there is a difference in microstructure along the SDAS. Robson and Prangnell [7] has looked at the effect of Zn, Cu and Mg on precipitation of $Al_3Zr$ dispersoids. However, the effect on the initial microsegregation of Zr which in turn affects the precipitation behavior has not been considered. In this study, we try to evaluate the microstructure both in the interdendritic regions and across the grains which is affected by variations in compositions.

For ease of homogenization, (i) the initial volume fractions of the interdendritic fraction should be minimum and (ii) the transformed S phase volume fractions should be low so that a homogenized structure with a uniform distribution of $Al_3Zr$ dispersoids and minimum S phase in α-Al is achieved in minimum time. For ease of extrusion which follows homogenization, (i) S phase in the homogenized alloy should be minimum (affects hot ductility) and (ii) there should be a high number density of fine nano-sized $Al_3Zr$ dispersoids across the grain (to inhibit recrystallization). Compositions which lead to these microstructural goals are investigated.

The current study is a comprehensive study of the effect of composition on microstructure after homogenization at two different length scales. A numerical model has been developed which couples microstructure development in the interdendritic regions involving transformation of η to S phase and their subsequent dissolution with the precipitation of $Al_3Zr$ dispersoids across the SDAS. The numerical model with its results and experimental validation can be found in greater detail in dissertation work by P. Priya[14]. Experiments have been performed to verify the microstructural evolution during homogenization for a solute rich and solute lean specimen. The consequences of the microstructure which varies with composition, on recrystallization and mechanical properties has also been discussed. Improved composition ranges for better extrudability with minimum recrystallization and better age-hardenability have been suggested.

## 2. Numerical Model

During homogenization of Al-Zn-Cu-Mg-Zr alloys, microstructural evolution takes place at two length scales:

(i) The η phase formed during solidification in the interdendritic regions near the grain boundaries transforms to the S phase followed by dissolution of the S phase.

(ii) The Zr microsegregated across the SDAS leads to precipitation of nano-sized coherent dispersoids of metastable $Al_3Zr$.

A numerical model has been developed which simulates both these microstructural changes using a cellular automaton finite volume model for the η to S phase transformation and a smaller scale finite difference model for $Al_3Zr$ precipitation. The models have been coupled together to predict temporal evolution of microstructure during homogenization. The model has been discussed in detail by Priya et al[15]. A review of the algorithm has been provided in Fig. 1. The initial microstructural domain represents a half SDAS with an initial microsegregation of the alloying elements based on Scheil type calculation as predicted by Thermo-Calc™. The halfgrain domain consists of 10 cells in 1D. Cell 1 is the interdendritic cell near the grain boundary where the η and S phases are present while cell 10 represents the center of the grain or the SDAS. While the CA-FV model is responsible for the η to S phase transformation in cell 1, the FD model simulates precipitation in each of the cells. The CA-FV model also calculates diffusion across all the cells during the homogenization holding time.

## 3. Experimental Procedure

To study the effect of composition on microstructure two separate samples with different compositions were prepared. The alloy of composition Al-6.2Zn-2.4Cu-2.3Mg-0.13Zr was first statically cast using a tilt pour vacuum induction furnace which was evacuated and back filled with Ar. Eutectic alloys of Al-33Cu, Al-36Mg and Al-88Zr were first cast from pure metals Al (99.99%), Zn (99.9999%), Cu (99.999%), Mg (99.8%), and Zr (99.8%). These were then added to prepare the desired composition. The molten metal was poured into Cu molds.

The statically cast alloy was then remelted and directionally solidified (DS) in alumina crucible by induction heating in a He atmosphere. The crucible was moved through the induction coils at a speed of 25 mm/hr. Due to segregation of the alloying elements, a solute lean top and a solute rich bottom of the directionally solidified sample was then sectioned to get two variants of compositions. It should be noted that Zr content was less in the bottom while it was more in the top sample due to a partition coefficient of 1.4 which is unlike other alloying elements which have a partition coefficient less than 1.

The DS top and bottom samples were then homogenized for a heating schedule of 420°C for 5 hrs followed by 480°C for 24-40 hrs in a box furnace. The microstructure of the coarse intermetallic particles and the nano-sized dispersoids were then characterized using an Olympus BX41M optical microscope (OM) and FEI XL40 field emission scanning electron microscope (FE-SEM) respectively. Graff-Sargent reagent (0.5vol.%HF, 15.5vol.%$HNO_3$, 84vol.%$H_2O$, and 3g$CrO_3$) was used for etching the optical samples. The $Al_3Zr$ dispersoids were quantitatively analyzed with ImageJ software using the Schwartz-Saltykov stereological method.

## 4. Results and Discussion

Effects of alloying elements Zn, Cu, Mg and Zr on the microstructure after homogenization has been numerically studied. The numerical microstructures after homogenization at 450°C for various compositions have been compared. The initial volume fraction of the η in the as-cast microstructure and initial microsegregation of the alloying elements has been predicted by Scheil type calculations using Thermo-Calc™. The baseline case was that of Al-6Zn-2Cu-2Mg-0.13Zr pertaining to AA7050 alloy. Zn/Mg ratios in the range of 1.5 to 6 and Zn+Cu+Mg in the range of 8 to 14 wt% have been investigated.

To study the effect of composition on microstructure and recrystallization behavior, two cases pertaining to the compositions corresponding to experimental directionally solidified top and bottom samples have also been studied. These samples were homogenized following a two-step process of 5hrs at 420°C and 40hrs at 480°C. The various compositions that have been studied numerically are provided in Table 1.

**Table 1**: The different numerical test cases run for different compositions. The bottom two rows also correspond to the experimental alloys.

| | **Zn** | **Cu** | **Mg** | **Zr** | **Zn/Mg** | **Zn+Cu+Mg** |
|---|---|---|---|---|---|---|
| **Zn1** | 4 | 2 | 2 | 0.13 | 2 | 8 |
| **Zn2/Cu2/Mg2** | 6 | 2 | 2 | 0.13 | 3 | 10 |
| **Zn3** | 8 | 2 | 2 | 0.13 | 4 | 12 |
| **Zn4** | 10 | 2 | 2 | 0.13 | 5 | 14 |
| **Cu1** | 6 | 1 | 2 | 0.13 | 3 | 9 |
| **Cu3** | 6 | 3 | 2 | 0.13 | 3 | 11 |
| **Cu4** | 6 | 4 | 2 | 0.13 | 3 | 12 |
| **Mg1** | 6 | 2 | 1 | 0.13 | 6 | 9 |
| **Mg3** | 6 | 2 | 3 | 0.13 | 2 | 11 |
| **Mg4** | 6 | 2 | 4 | 0.13 | 1.5 | 12 |
| **Zr1** | 6 | 2 | 2 | 0.05 | 3 | 10 |
| **Zr2** | 6 | 2 | 2 | 0.10 | 3 | 10 |
| **Zr3** | 6 | 2 | 2 | 0.15 | 3 | 10 |
| **Zr4** | 6 | 2 | 2 | 0.20 | 3 | 10 |
| **DS_Top** | 6.6 | 3.4 | 4.3 | 0.03 | 1.5 | 14.3 |
| **DS_Bottom** | 5.1 | 1.4 | 2.2 | 0.11 | 2.3 | 8.7 |

### *4.1. Effect of composition on interdendritic phases*

Composition of the alloys affects the phases formed during solidification as well as the phase transformations during homogenization. It also affects the phases present post-homogenization downstream processing: extrusion and age-hardening which will not be discussed in the current work.

### 4.1.1. *Effect on as-cast and homogenized microstructure*

The microstructural evolution during homogenization for different compositions can be compared only when we begin with the right volume fractions of the interdendritic phases present in the as-cast microstructure which in turn depends on the compositions. Scheil type solidification calculations were performed using Thermo-Calc$^{TM}$ which disregards the diffusion in the solid phase and diffusion in the liquid phase tends to infinity considered to be a good approximation for practical cases. The compositions not only affect the as-cast phases but also the microsegregation of various alloying elements which has a considerable effect on the segregation of Zr and $Al_3Zr$ precipitation which will be discussed in the later section.

Varying amounts of Zn, Cu and Mg in the alloy leads to varying amounts of T phase, V phase and the S phase in the as-cast microstructure. While T phase is a solution of the $MgZn_2$ phase and the η phase represented as $(Al,Cu,Zn)_{49}Mg_{32}$, the V phase also known as the Z phase is a solution of $Mg_2Zn_{11}$ and $Al_5Cu_6Mg_2$ with varying Cu and Al solubilities. S phase is mostly stoichiometric represented as $Al_2CuMg$. For the sake of numerical calculations, the T and V phases are treated as a single solid solution which transform to the S phase which has also been experimentally observed [16], [17].

The effect of Zn, Cu and Mg on the as-cast interdendritic phases calculated using ThermoCalc$^{TM}$ using the TCAL1 thermodynamic database [15], is shown in Fig. 2. The homogenization kinetics showing varying time lengths of Stage I pertaining to T+V to S phase transformation, Stage II pertaining to slower growth rate of S till dissolution of T+V and Stage III pertains to dissolution of S to its equilibrium volume fraction, is shown in Fig. 3. Increase in Zn leads to increase in the Zn-rich T and V phases which transform to S phase during homogenization, as seen in Figure 3(b). Increase in Zn levels lead to longer Stage II transformations which involves complete dissolution of the as-cast T+V phases. Zinc levels more than 8% lead to larger volume fractions of T+V in the as-cast material which are difficult to dissolve as seen in Figure 3(a). It is interesting to note no linearity in volume fraction change at higher solute contents, indicating the transformations may be diffusion controlled. Decreasing the Zn levels on the other hand, does not help the homogenization kinetics either, as the fraction of the S phase increases due to increased Cu and Mg contents. This S phase takes more time to dissolve leading to longer Stage III which continues till the equilibrium volume fraction of S is reached.

Increase in Cu leads to decrease in the Zn-rich T phase, and an increase in the Cu-rich S phase and V phase, which has increased Cu solubility, in the as-cast microstructure, as seen in Figure 2. The increased T+V volumes lead to longer Stage I transformations to S phase during homogenization, resulting in larger volume fractions of S phase as seen in Figure 5.2(d). This S phase needs to be minimized leading to longer Stage III transformations. Low Cu compositions of ~1%, on the other hand, leads to a short Stage II and even shorter Stage III as the volume fraction of S phase formed is low which easy dissolves to give a S phase free microstructure which is desirable. However, intermediate amounts of Cu is needed in the alloy for strength and ductility[18].

Increase in Mg content leads to an increase in the Mg-rich T and S phases and a decrease in the Zn-rich V phase in the as-cast microstructure as seen in Figure 2(c). The S phase is seen to decrease after 3% Mg as it is replaced by T phase richer in Mg. An interesting thing to note during homogenization of Mg rich alloys is an initial increase in T phase and decrease in S phase owing to microsegregation of Mg which leads to a very high concentration of Mg near grain boundaries which favors a reversion of S phase to T phase (Figure 3(e) and (f)). Increasing the Mg content leads to an increase in the Stage I and Stage II transformations when the T phase transforms to S phase. Mg of 4% leads to a prolonged Stage II without a Stage I, which extends to more than 50 hours as seen in Figure 3(e). There is no linearity in the volume fraction change indicating diffusion-controlled transformations for high Mg content of 4%. The amount of transformed S is minimum for 1% Mg (minimum Mg) which has a short Stage II and Stage III leading to complete dissolution of S phase during homogenization as seen in Figure 3(f). Magnesium however, is a necessary alloying element needed for the strengthening η' precipitates.

### *4.1.2. Effect on phase diagrams*

Table 2 summarizes the effect of composition on as-cast microstructure and time taken for homogenization. The effect of composition on microstructural evolution witnessed in this study is because these alloying elements affect the phase diagrams for quaternary Al-Zn-Cu-Mg system. The phase diagram information is incorporated in the numerical model from Thermo-Calc$^{TM}$ using the TCAL1 database. An older version for these phase diagrams is available from Stawbridge et al. [19] An updated version from Thermo-Calc$^{TM}$ is presented in Fig. 4 for compositions of 4%, 6%, 8% and 10% Zn at 450$^{o}$C.

**Table 1**: Summary of effect of composition on as-cast microstructure and homogenization time

| | **As-cast** | **Homogenized** |
|---|---|---|
| Higher Zn | More T, More V | More time for stage II and stage III |
| Higher Cu | More S and V, Less T | Way more time for stage III |
| Higher Mg | More T and S, Less V | More time for stage I and lesser time for stage III |

With increase in Zn content the number of stable phases in the composition range investigated increases. At low Zn content, Mg-rich T is stable at low Cu contents and Cu-rich ϴ is stable at low Mg contents as seen in Figure 4(a). At intermediate compositions of Cu and Mg the S phase stabilizes. With increase in Zn, V phase with high Cu solubility stabilizes for higher Cu contents as seen in Figure 4(b). For still higher Zn of 6% the η phase appears as seen in Figure 4(c). For Zn as high as 10%, the Zn-rich V phase regions expand. The as-cast and homogenized microstructures, as expected, show trends similar to the phase diagrams.

## 4.2. *Effect of composition on $Al_3Zr$ dispersoids*

Composition not only affects the microstructure in the interdendritic regions near the grain boundaries but also affects the microstructure within the grains. This happens due to difference in Zr concentrations across the grains brought in by microsegregation during casting. Zirconium has a very low diffusivity owing to which it is not "homogenized" even after prolonged durations of holding at homogenization temperature ranges. Presence of Zr above the solubility limits during homogenization causes the precipitation of $Al_3Zr$ dispersoids whose number density and radius and hence the microstructure across the grains depends on the (a) initial Zr concentration; (b) solubility limits (c) nucleation and growth rates all of which are affected by composition and temperature. These in turn affect the homogenized microstructure which otherwise needs a uniform distribution of fine $Al_3Zr$ dispersoids.

### 4.2.1. *Effect on Zr micro-segregation*

As the difference in microstructure across the grains occurs due to the microsegregation of Zr that was caused during solidification, the microsegregation in the initial microstructure to start with for different compositions should match as-cast microsegregations. The microsegregations have been predicted from Scheil type calculations using Thermo-Calc$^{TM}$. The solidification of Al-Zn-Cu-Mg starts with crystallization of equilibrium $L1_2/DO_{23}$ $Al_3Zr$ in the matrix followed by crystallization of the fcc α-Al phase. So, some Zr is lost in precipitating out these primary peritectic precipitates.

The equilibrium mass fraction of $DO_{23}$ $Al_3Zr$ (which is predicted by Thermo-Calc$^{TM}$) for various compositions investigated as seen in Fig. 5(a), we find higher alloying contents lead to higher equilibrium mass fraction of $DO_{23}$ $Al_3Zr$ leading to less of it to remain in the matrix. This translates to the decreasing amount of Zr along the SDAS or in the grain for example at higher Mg content as seen in Fig. 5(b). The same has been observed for other alloying elements too.

#### *4.2.2. Effect on thermodynamics and kinetics of precipitation*

The solubility limits of Zr in the fcc α-Al matrix depends on the composition of the alloy. The change in fcc α-Al phase solvus with increasing Mg concentration is shown in Fig. 6(a). It is seen that the solubility of Zr decreases with increase in Mg content of the alloy leading to higher supersaturation and higher driving force for $Al_3Zr$ nucleation and growth. The same trend is true for increasing Cu and Zn compositions. However, Mg has a greater influence on solubility limits as compared to Zn or Cu.

The nucleation and growth rates for various compositions in the order of increasing Mg content for various temperatures are shown in Fig. 6(b) and (c) respectively. Both the nucleation and growth rates increase with increasing temperatures owing to increased diffusivity at higher temperatures. The nucleation and growth rates are same for all compositions till a temperature of 450°C after which increased Mg compositions lead to increase in both nucleation and growth rates as seen in Fig. 6(b) and (c). To study the effect of composition on dispersoid precipitation (dispersoid number density and size) at two temperatures of 450°C and 470°C have been further studied pertaining to the regimes with no difference and some difference in nucleation and growth rates as shown in Fig. 7.

#### *4.2.3. Effect on dispersoid precipitation*

The effect of varying amounts of Zn, Cu, Mg and Zr on the dispersoid microstructure across the grains about the baseline case of Al-6Zn-2Cu-2Mg-0.13Zr has been shown in Fig. 7. The number densities and mean radii of the dispersoids at temperature of 450°C on homogenization for 30 hrs ((a),(c),(e)) and 470°C on homogenization for 5 hrs ((b),(d),(f)) have been compared.

Both the number densities and mean radii decrease with increasing amounts of Zn, Cu and Mg, while both increase with increasing amount of Zr. This is true for both the temperatures, 450°C pertaining to regime with no difference in nucleation and growth rates and 470°C pertaining to the regime with difference in nucleation and growth rates as can be seen in Fig. 6(c) and (d).

The trend observed in this study is reverse to the trend observed by Robson et al.[7] who did not consider the initial microsegregation of Zr in the as-cast microstructure into account. Although the solubility of Zr is decreased with increasing alloy content leading to increase in nucleation and growth rates, the availability of lesser amount of Zr along the grains or SDAS with increasing alloying content restricts both the number density and mean radii. Therefore, the initial microsegregation is very crucial in determining the effect of composition on dispersoid precipitation.

The effect of Zr content on microstructure on homogenization at the two temperatures can be seen in Fig. 7(g) and (h) respectively. It can be seen that at 450°C, the mean radius of the dispersoids see a sharp increase after 0.15% Zr which is not desirable as a very fine distribution of dispersoids is required to pin grain boundaries during recrystallization [20]. Also, both the number density and mean radius seem to saturate after 0.15% leading to no gain in benefits of adding more Zr above 0.15% Zr.

### *4.3. Experimental validation of the compositional variations*

To study the effect of composition on the evolution of microstructure during homogenization, two samples, one from the top and other from the bottom part of the directionally solidified by vertical zone melting was chosen. Due to segregation, while the bottom is lean in Zn, Cu and Mg and rich in Zr while the top is rich in Zn, Cu and Mg and lean in Zr. This is because while Zn, Cu and Mg have a partition coefficient less than 1, Zr has it greater than 1. The exact composition of the samples from EDX/OES methods is Al-5.1Zn-1.4Cu-2.2Mg-0.11Zr for the bottom sample and Al-6.6Zn-3.4Cu-4.3Mg-0.03Zr for the top sample.

The DS top and bottom samples were homogenized for 5hrs at 420°C and for 24 to 40 hrs at 480°C. The microstructure both in the interdendritic regions and across the SDAS was characterized experimentally using optical microscopy, SEM and analyzed using ImageJ software. The area fraction of the interdendritic phases was measured from the optical micrographs which is equal to the volume fraction of the phases by stereology. The number density of dispersoids per unit area was converted into number density per unit volume by the Schwartz Saltykov method. The mean radius was eventually calculated.

Numerical test cases were set up pertaining to the compositions of the two samples and were run for the homogenization schedule that had been experimentally performed. The initial microstructure was chosen as predicted by Thermo-Calc$^{TM}$. The comparison of the microstructure at the two length scales is provided in Table 3. The results for dispersoids are within the experimental errors. The average experimental size of the dispersoids is higher because of the larger primary dispersoids which might have been formed during zone melting. Also, images of dispersoid rich zones have been analyzed which might lead to overprediction of the number densities. The model predicted melting for the DS top sample while it predicted full dissolution of the interdendritic particles for the bottom sample. The discrepancy for the top sample may be because of the variation of melting point of the remnant S phase in the sample which seems to be higher for the sample than that predicted by Thermo-Calc$^{TM}$ as no melting was observed in the sample. The microstructures after homogenization for both the samples can be seen in Figure 8.

**Table 3**: Comparison of the predicted and experimentally measured microstructure of the DS Top and DS Bottom samples after homogenization for 5hrs at 420°C and 24hrs for interdendritic phase 40hrs for dispersoids at 480°C.

| | **DS Top** | | **DS Bottom** | |
|---|---|---|---|---|
| | **Experimental** | **Numerical** | **Experimental** | **Numerical** |
| Interdendritic phases | | | | |
| Volume fraction (%) | 1.5±0.89 | Melting | 1.0±0.97 | 0 |
| $Al_3Zr$ dispersoids | | | | |
| Number density (/µm$^3$) | 725 | 528 | 826 | 583 |
| Mean diameter (nm) | 28.5±12.4 | 15.2 | 37.1±20.0 | 17.7 |

The recrystallization behavior of these two samples have been compared by Yiwei et al. [21] and in spite of the higher Zr in the bottom sample, it is found to be less resistant to recrystallization. This can be attributed to the coarse dispersoids found for this sample which has also been predicted by the model.

### *4.4. Improved composition ranges*

Increasing Zn increases the time needed for homogenization. However, Zn above 6 wt% minimizes S phase considerably but leads to other phases like T and V in the as-cast microstructure, which need to be dissolved before extrusion and age-hardening by homogenization at higher temperatures. Having Zn higher than 8% leads to very high alloying element content increasing the quench sensitivity of the alloy requiring high cooling rates to subdue precipitation which can induce residual stress and is also difficult to attain for thick forgings. So, a Zn content of 6-8% with Zn/Mg ratio of 3-4 is desirable.

High Cu or Mg increases the S phase that remains after homogenization which is difficult to dissolve. Low Cu and Mg (~1%) leads to easy homogenization with no detrimental interdendritic particles which is desirable. However, they are both needed for good mechanical properties of the alloy. The desirable Cu and Mg content is in the range of 1-2% with Mg:Cu of 1-2.

The composition of Zn, Cu and Mg affect the amount of primary $Al_3Zr$ precipitated during solidification, leaving remaining Zr available for precipitation of fine dispersoids during homogenization. These precipitates formed during solidification are coarse and incoherent, and hence undesirable. In general, increase in alloying content increases the tendency for primary precipitation during solidification leading to fewer fine coherent dispersoids at high alloy contents. This is however the reverse, if cooling during solidification exceeds a critical cooling rate leading to less or no precipitation of $Al_3Zr$. Alloying elements decrease the solid solubility of Zr leading to higher volume fractions of metastable $Al_3Zr$ which is desirable. But this advantage can be taken only when we can prevent primary $Al_3Zr$ from precipitation during solidification. No significant gain in number density or mean radius observed above 0.15% Zr at both temperatures investigated in the study. Hence a Zr content of 0.1-0.15% is the optimum range to attain fine distribution of numerous metastable coherent $Al_3Zr$ dispersoids.

## 5. Conclusions

The effect of composition on microstructural evolution during homogenization of Al-Zn-Cu-Mg-Zr alloys has been investigated. A comprehensive model that was developed to simulate microstructural changes of undesirable interdendritic particles and desired fine dispersoids across the grains has been used for the study. The results have been compared to the experimental microstructures from two samples of a directionally solidified alloy sample.

The composition affects the volume fraction of secondary particles and microsegregation during solidification which has a profound effect on the microstructure during homogenization and subsequent processing. Higher Zn, Cu and Mg contents lead to higher amounts of interdendritic particles and hence require more time to homogenize, in general. Higher alloying content also leads to increased quench sensitivity. Higher alloy content also leads to decrease in solid solubility of Zr leading to higher driving force for nucleation of dispersoids. This is, however, possible only when the cooling rate during solidification is fast enough to prevent primary $Al_3Zr$ from nucleating. These primary precipitates are coarse and incoherent and decrease the amount of Zr needed for precipitation of dispersoids which is undesirable. For solidification under normal conditions higher alloying content leads to lower number densities for dispersoids.

Based on the study an improved composition range of 6-8%Zn, 1-2%Cu, 1-2%Mg and 0.1-0.15%Zr has been suggested.

## Acknowledgement

The authors acknowledge financial support for this research from Shandong Nanshan Aluminum Co., Beijing Nanshan Institute of Aeronautical Materials.

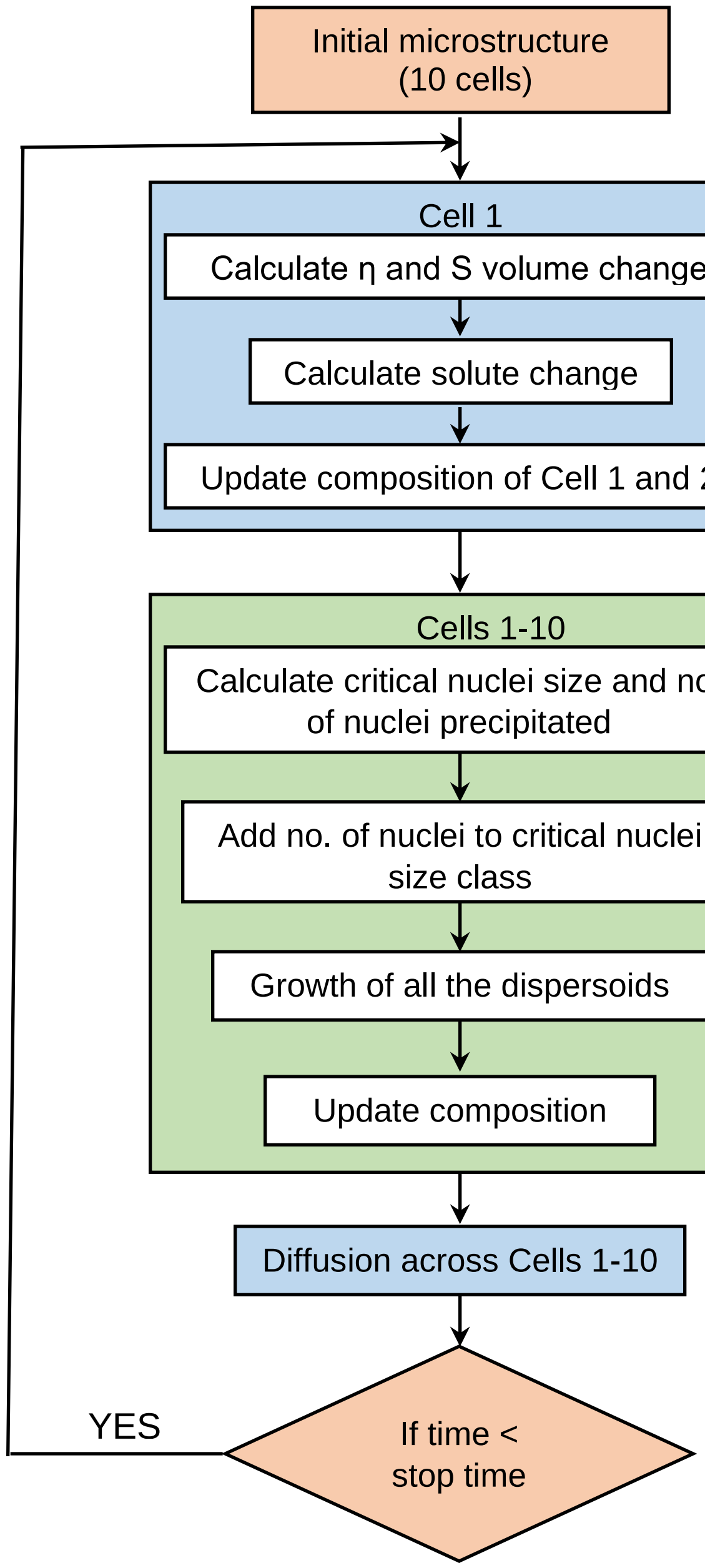

**Figure 1**: The flow chart showing the algorithm to simulate microstructural evolution in the interdendritic region and across the SDAS.

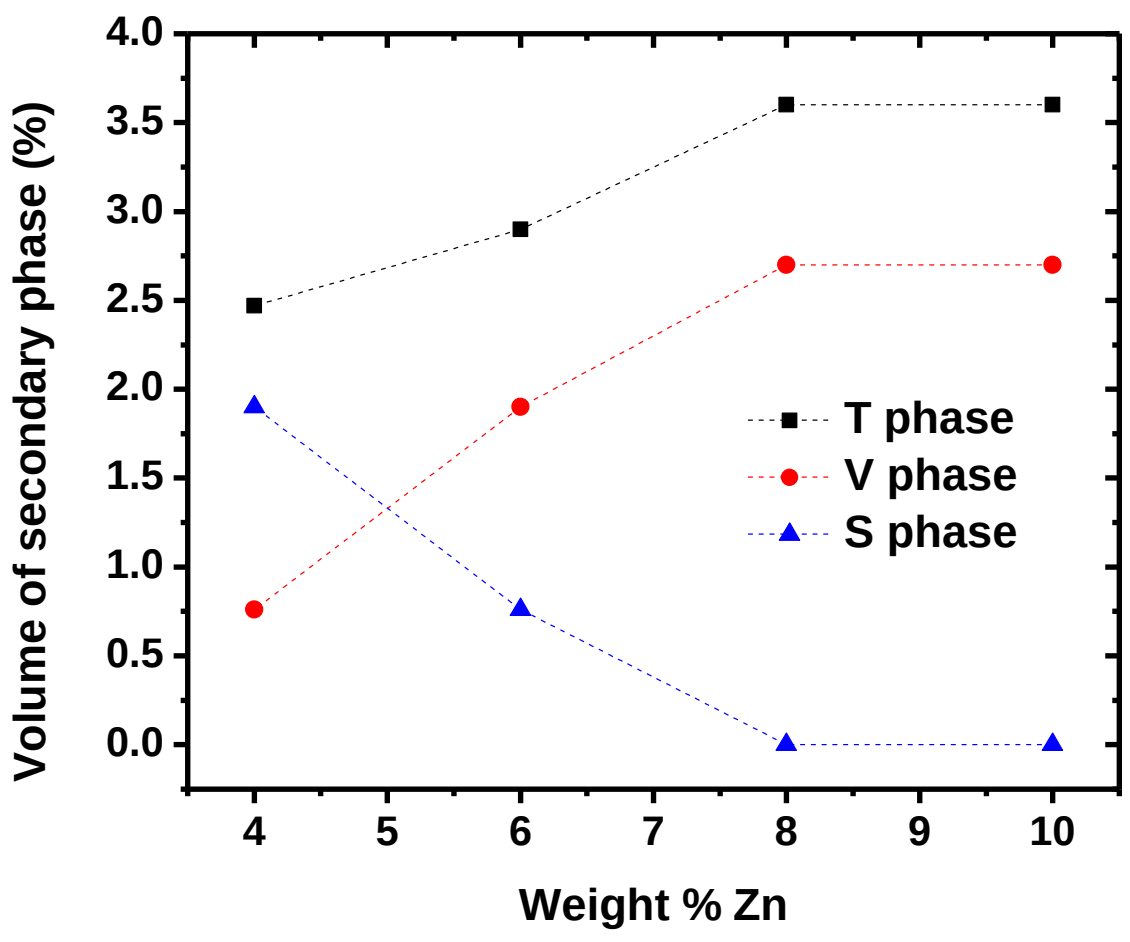

(a)

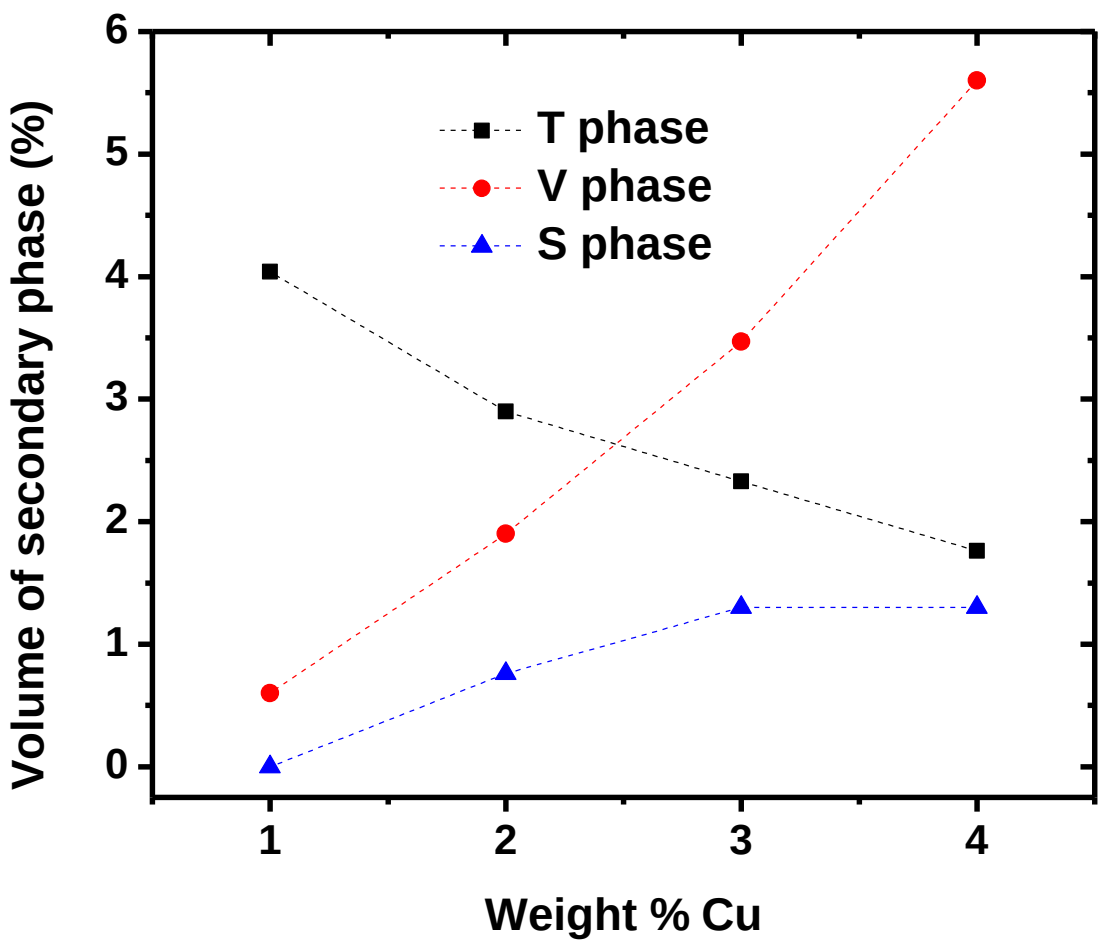

(b)

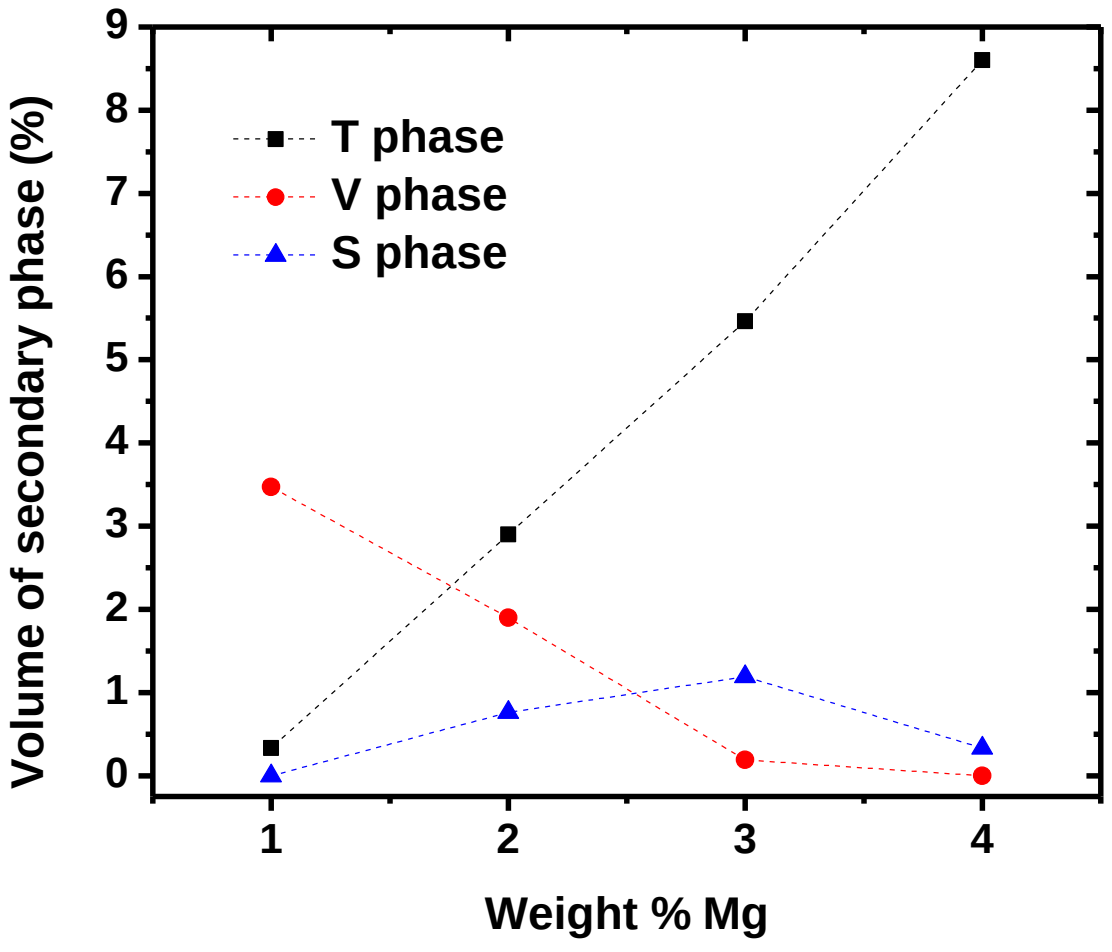

(c)

**Figure 2**: Effect of varying amounts of (a) Zn, (b) Cu and (c) Mg on the initial volume fraction of interdendritic particles in the as-cast state predicted by Thermo-Calc™.

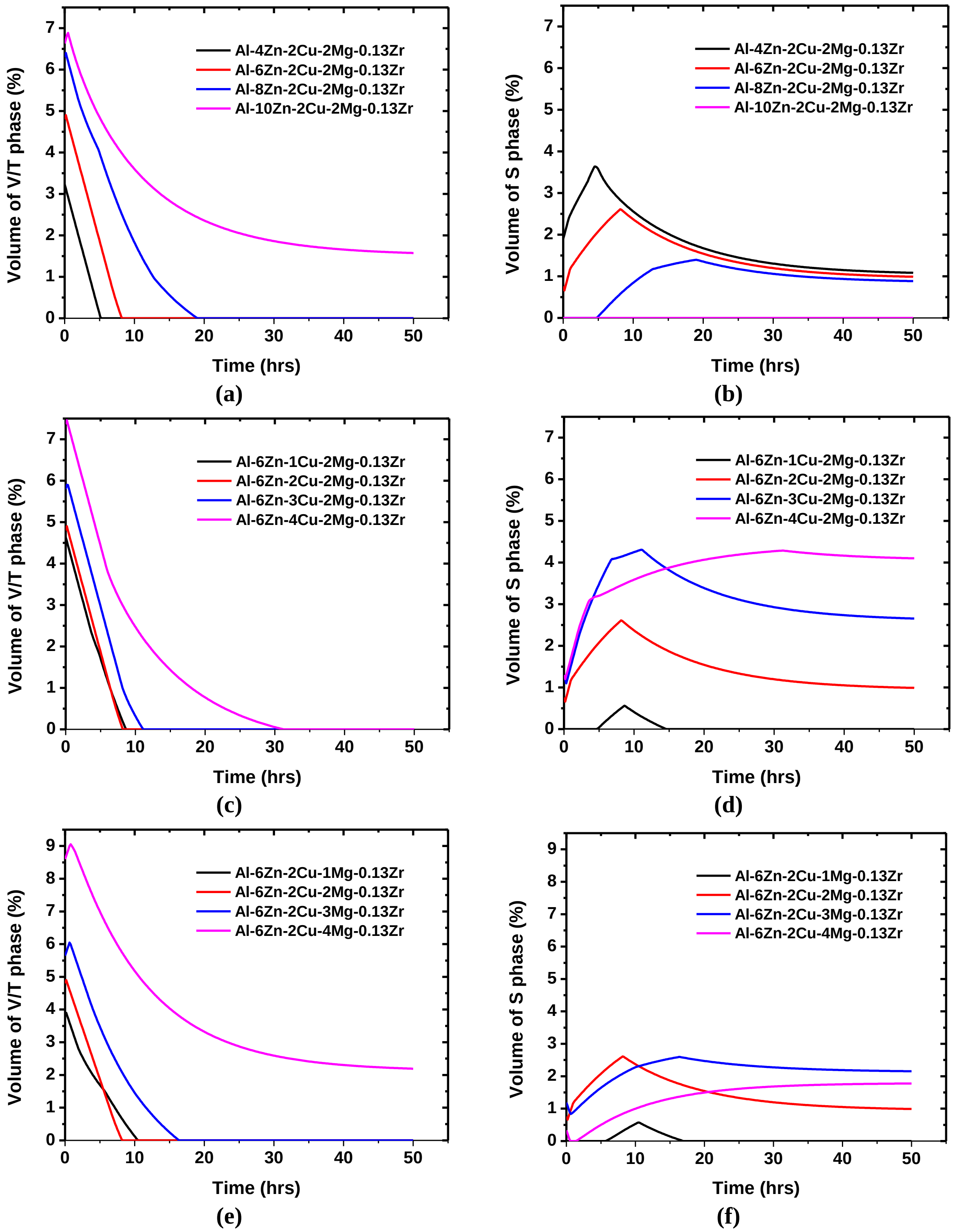

**Figure 3**: Effect of alloying elements (a),(b) Zn; (c),(d) Cu and (e),(f) Mg on evolution of T/V and S phases respectively during homogenization at 450°C.

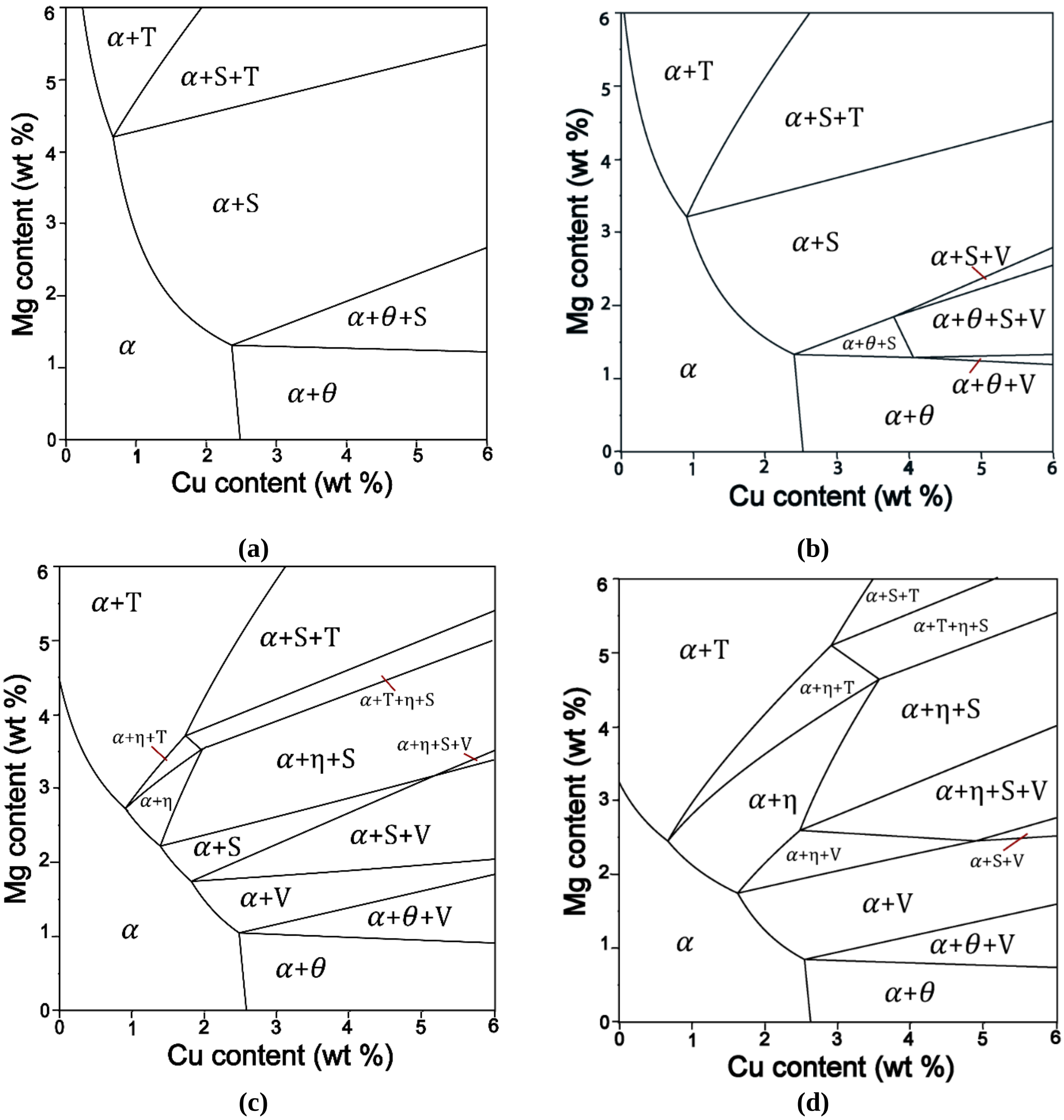

**Figure 4:** Effect of Cu and Mg on phase diagrams for (a) 4%, (b) 6%, (c) 8%, (d) 10% Zn at 450°C

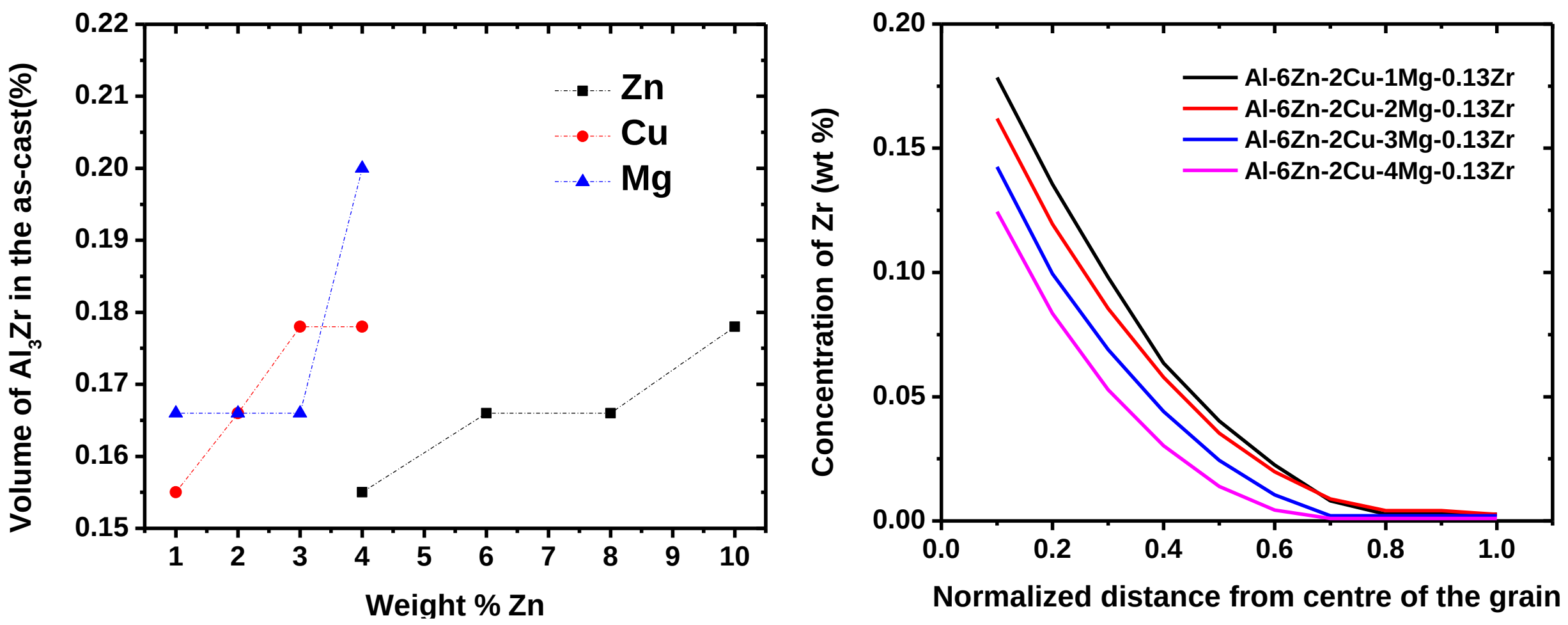

(a) (b)

**Figure 5**: (a) Variation of volume fraction of $Al_3Zr$ in the as-cast state with composition for the baseline case of Al-6Zn-2Cu-2Mg-0.13Zr predicted by Thermo-Calc™ (b) Composition of Zr across the SDAS for variation of Mg

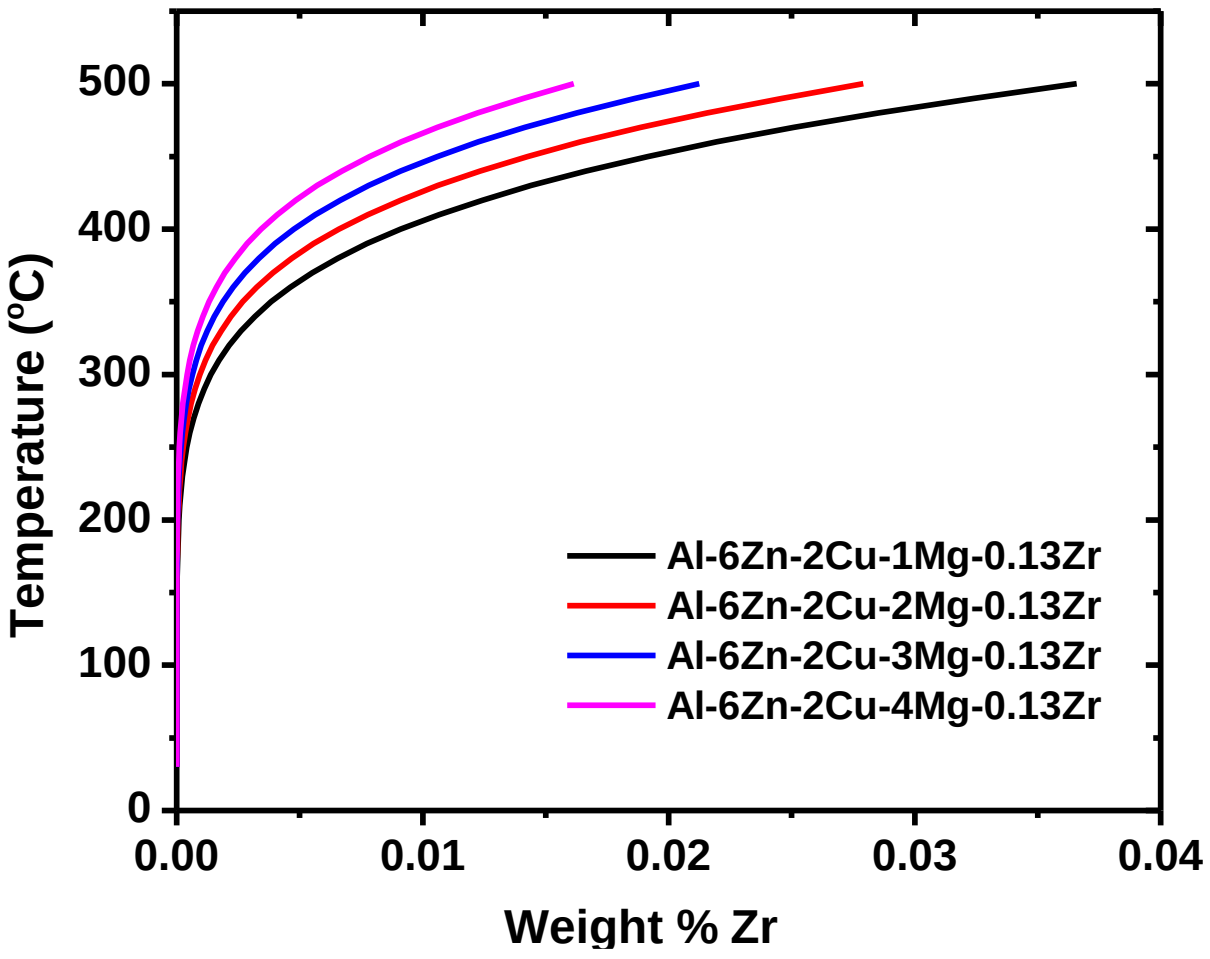

(a)

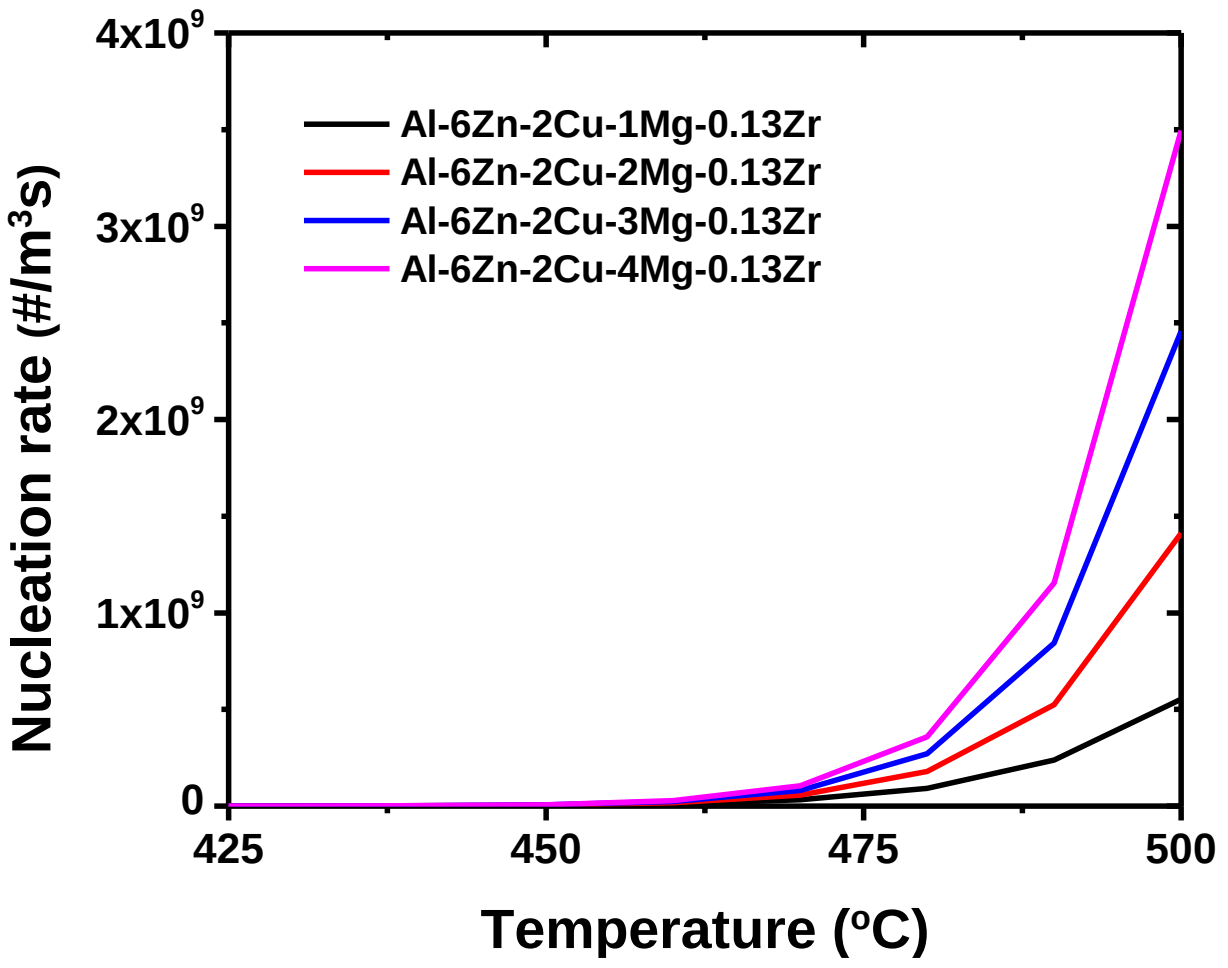

(b)

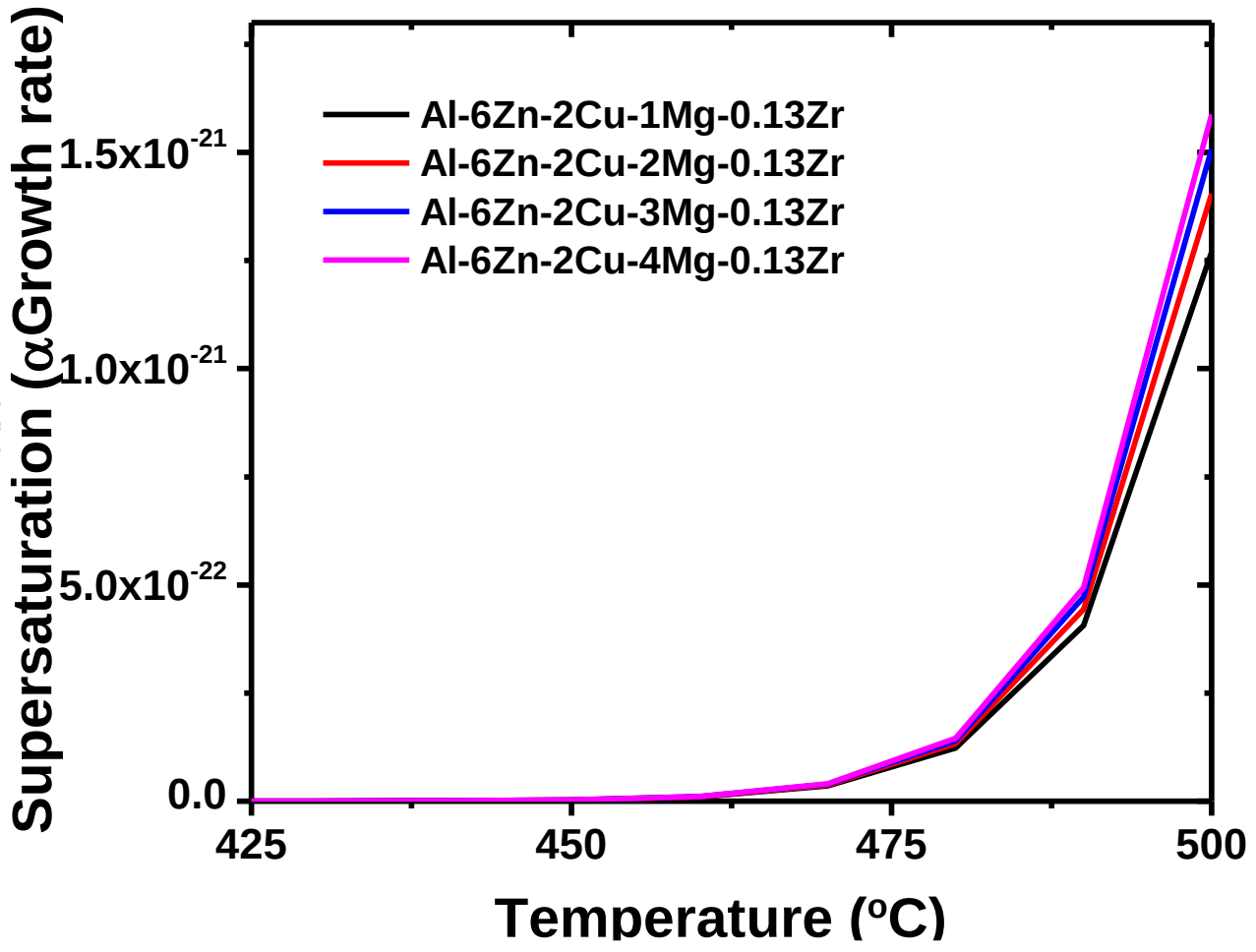

(c)

**Figure 6**: Variation of (a) supersaturation, (b) nucleation, (c) growth rates for varying amounts of Mg for the baseline case of Al-6Zn-2Cu-2Mg-0.13Zr

| 450°C for 30 hrs | 470°C for 5 hrs |
| --- | --- |

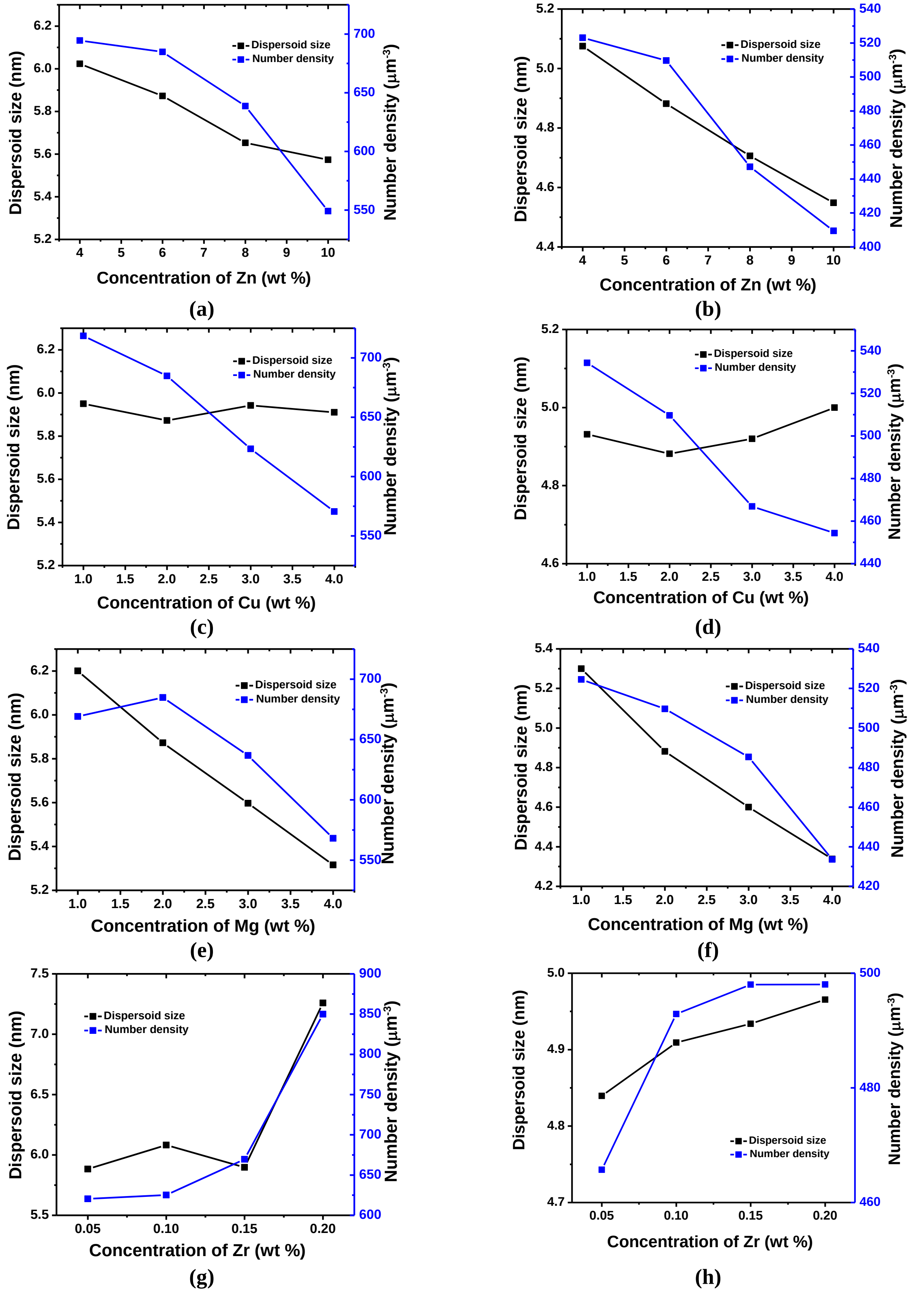

**Figure 7**: Effect of composition on number density and mean radius of the dispersoids (a),(b) Zn; (c),(d) Cu; (e), (f) Mg; (g),(h) Zr for homogenization at 450°C for 30 hrs and 470°C for 5 hrs respectively.

| DS Top | DS Bottom |
|---|---|

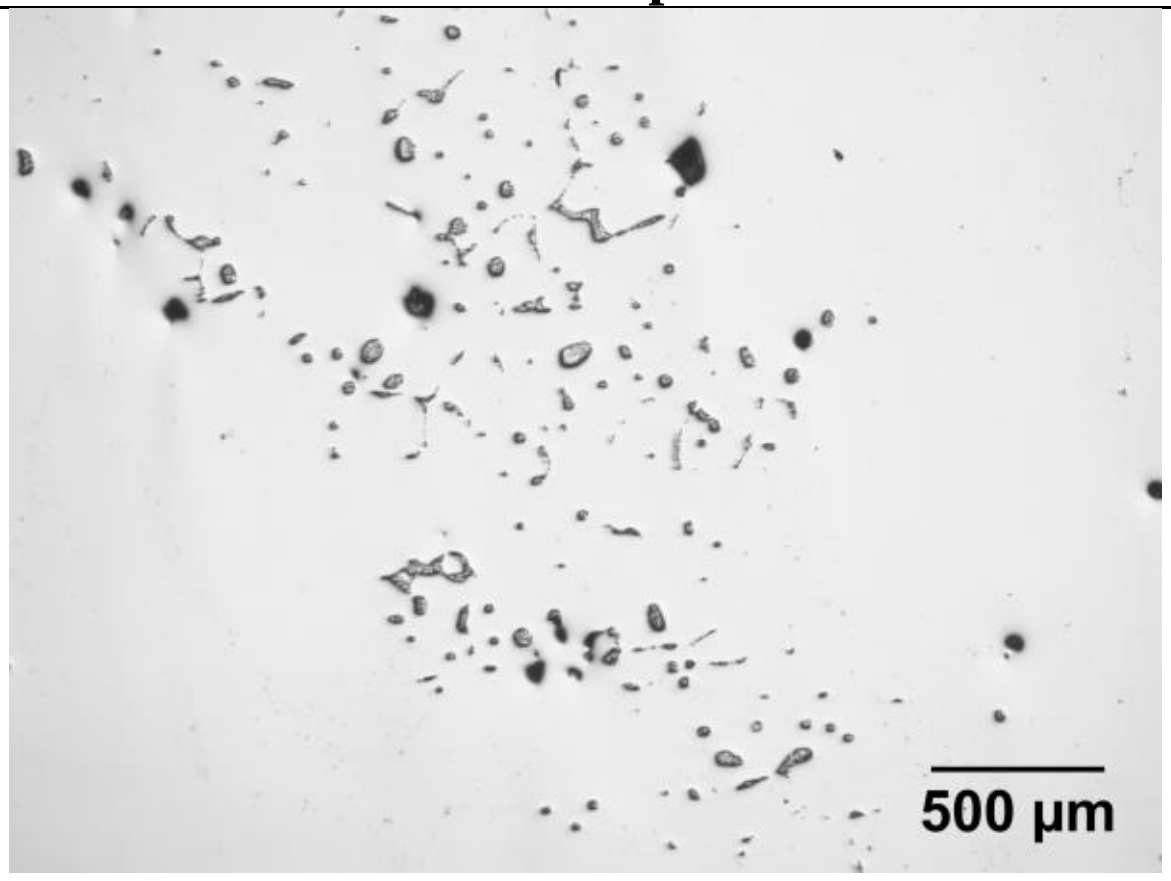

(a)

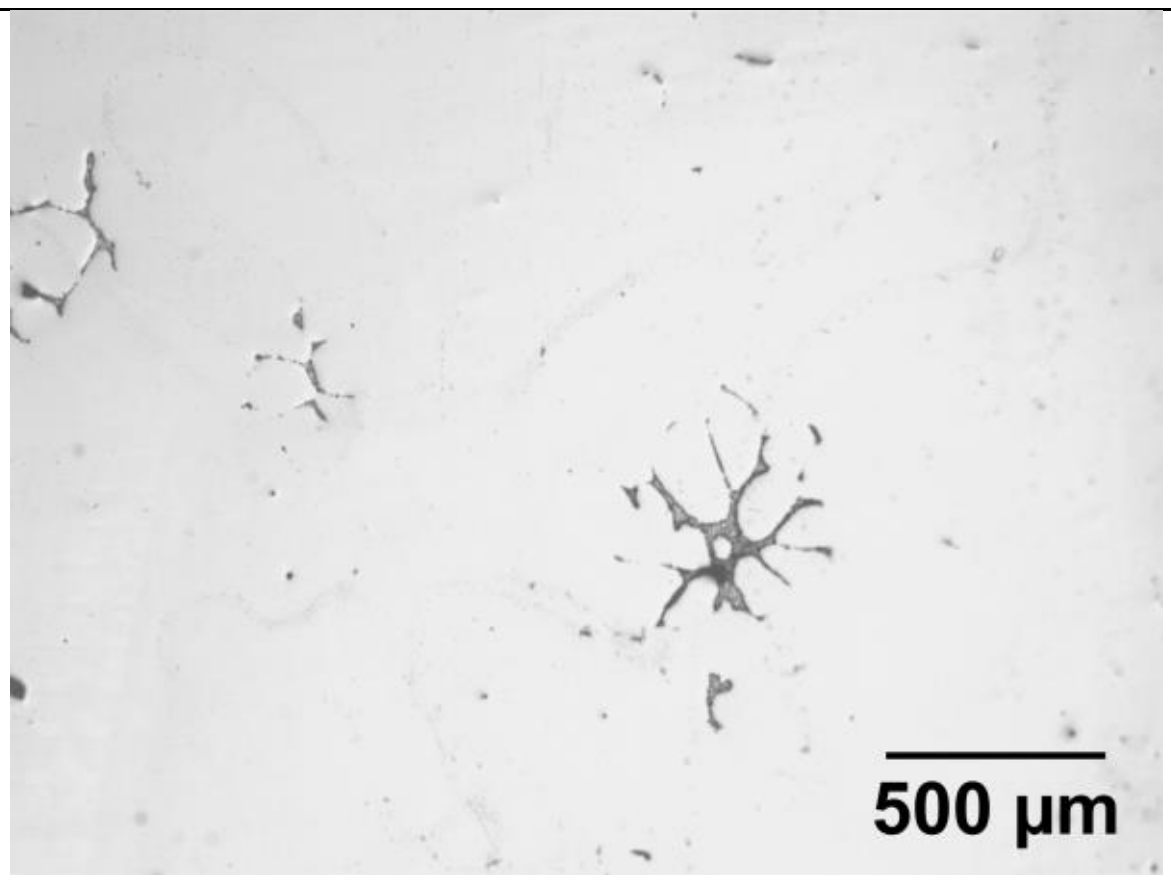

(b)

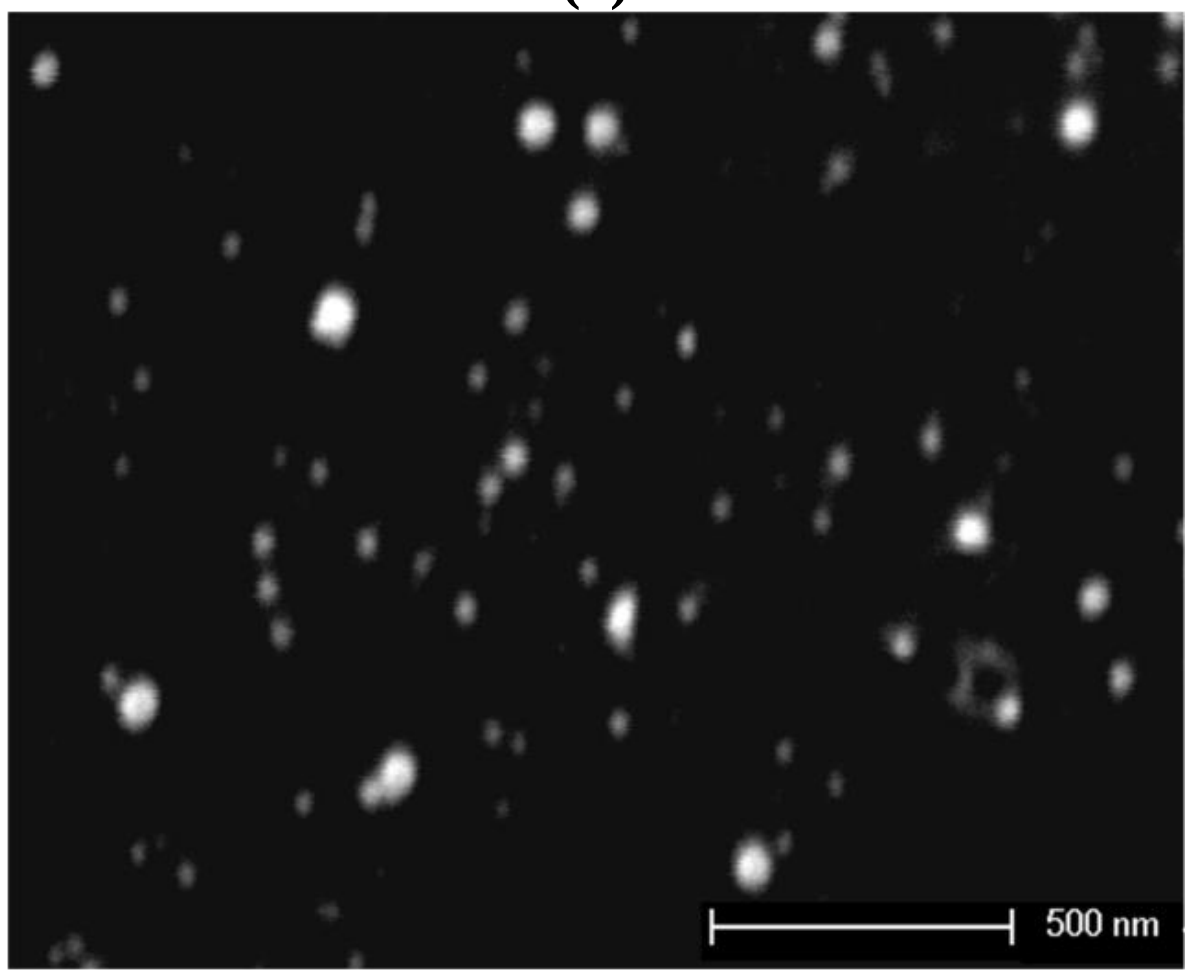

(c)

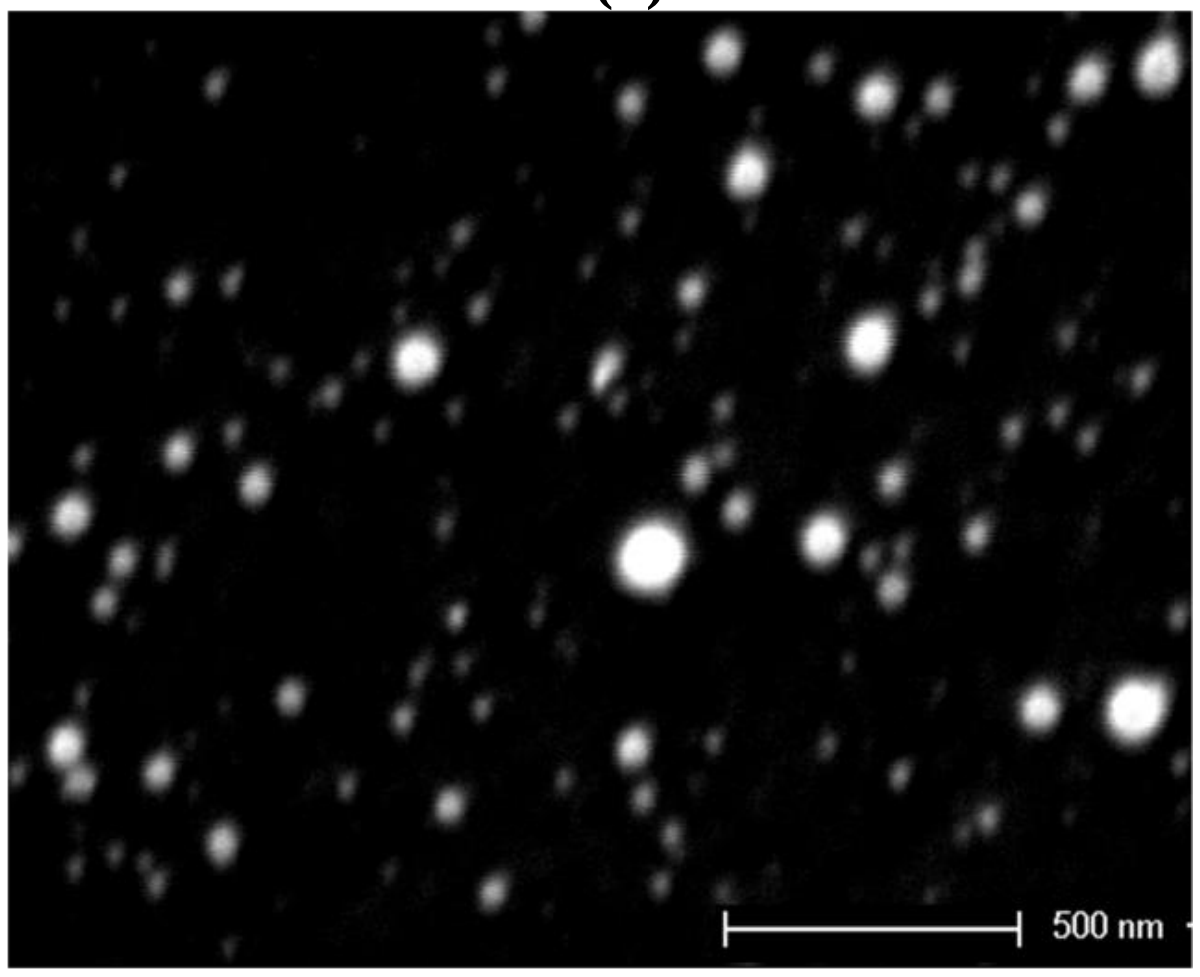

(d)

**Figure 8**: The interdendritic phases in the (a) DS top and (b) DS bottom samples after homogenization for 420°C for 5hrs + 480°C for 24hrs. The nano-sized dispersoids in the (c) DS top and (d) DS bottom samples after homogenization for 420°C 5hrs + 480°C for 40hrs.